\documentclass[11pt]{article}

\usepackage[utf8]{inputenc}
\usepackage{amsmath, amssymb, amsthm}
\usepackage[letterpaper, margin=1in]{geometry}
\usepackage{booktabs}
\usepackage{graphicx, psfrag, epsf}
\usepackage{enumerate}
\usepackage{enumitem}
\usepackage{amsfonts}
\usepackage{verbatim}
\usepackage{xcolor}
\usepackage{epstopdf}
\usepackage{caption, subcaption}
\usepackage{algorithm}
\usepackage{algpseudocode}
\usepackage[detect-all]{siunitx}
\usepackage{tabu}
\usepackage{arydshln}
\usepackage{eufrak}
\usepackage{mathtools}
\usepackage{placeins}
\usepackage{multirow}
\usepackage{rotating}
\usepackage{afterpage}
\usepackage{titlesec}

\DeclareMathOperator*{\argmax}{arg\,max}

\newcommand{\bbet}{\boldsymbol{\beta}}

\newcommand{\bu}{\mathbf{u}}
\newcommand{\bv}{\mathbf{v}}
\newcommand{\bx}{\mathbf{x}}
\newcommand{\bX}{\mathbf{X}}
\newcommand{\by}{\mathbf{y}}

\newcolumntype{"}{@{\hskip\tabcolsep\vrule width 1pt\hskip\tabcolsep}}
\usepackage{tikz}
\usetikzlibrary{shapes.geometric, arrows.meta, positioning, fit, backgrounds}

\definecolor{algblue}{RGB}{0, 80, 160}     
\definecolor{algcomment}{RGB}{0, 120, 80}  

\renewcommand{\algorithmicrequire}{\textbf{\textcolor{algblue}{Input:}}}
\renewcommand{\algorithmicensure}{\textbf{\textcolor{algblue}{Output:}}}

\algrenewcommand\algorithmicrequire{\hspace*{\algorithmicindent}\textbf{Input:}}
\algrenewcommand{\algorithmicensure}{\hspace*{\algorithmicindent}\textbf{Initialize:}}
\algdef{SE}[DOWHILE]{Do}{doWhile}{\algorithmicdo}[1]{\algorithmicwhile\ #1}%

\newcounter{algphase}
\newcommand{\algphase}[1]{%
	\refstepcounter{algphase}%
	\State \textcolor{algblue}{\textbf{--- Phase \thealgphase: #1 ---}}%
}

\usepackage{setspace}
\usepackage[authoryear, round]{natbib} 

\usepackage[colorlinks=true, citecolor=blue, linkcolor=blue, urlcolor=blue]{hyperref}

\title{\textbf{Fast and Scalable Cellwise-Robust Ensembles \\ for High-Dimensional Data \vspace{0.75cm}}}

\author{
	\Large \textbf{Anthony Christidis} \\
	Department of Statistics, University of British Columbia\\
	Department of Biomedical Informatics, Harvard Medical School \\[1.5em]
	\Large \textbf{Jeyshinee Pyneeandee} \\
	Department of Statistics, University of British Columbia \\[1.5em]
	\Large \textbf{Gabriela Cohen Freue} \\
	Department of Statistics, University of British Columbia
}

\date{\today} 

\begin{document} 
	
	\maketitle
	
	\begin{abstract}
		Variable selection and ensemble methods are central to high-dimensional modelling, enabling the identification of relevant predictors and the construction of stable predictive signals through aggregation across multiple models. However, in practice, high-dimensional data are often affected by cellwise contamination, in which individual cells of the data matrix deviate from the underlying multivariate structure without necessarily making the corresponding observation outlying. This type of contamination can easily propagate throughout many observations, compromising variable selection procedures and ensemble methods, including robust methods designed for contamination affecting entire observations (casewise contamination). To address this limitation, we propose the Fast and Scalable Cellwise-Robust Ensemble (FSCRE) algorithm. FSCRE dynamically partitions predictors into disjoint sub-models using a competitive proposer--arbiter architecture operating in a robust correlation framework. Through extensive simulations and a bioinformatics application, we demonstrate FSCRE's competitive performance in variable selection precision, recall, and predictive accuracy across various contamination scenarios, all while maintaining high computational efficiency in high-dimensional settings. This work provides a unified framework connecting cellwise-robust estimation with high-performance ensemble learning, with an implementation available on CRAN.
	\end{abstract}
	
	
	\newpage
	
	\section{Introduction}\label{sec:introduction}
	
	High-dimensional data, where the number of predictors $p$ far exceeds the number of observations $n$, are now ubiquitous across scientific domains, from genomics and proteomics to finance and neuroimaging. A primary goal in analyzing such data is variable selection: identifying a sparse subset of predictors that are genuinely associated with a response variable. In practice, however, real-world datasets often contain atypical or corrupted values, arising from measurement errors, data integration issues, or rare but influential events that may affect entire observations (casewise contamination) or individual entries of the data matrix (cellwise contamination). While numerous methods have been developed for variable selection, their fragility to such contamination remains a fundamental challenge. Classical methods, and even robust approaches designed for casewise contamination, can be highly sensitive to cellwise outliers, leading to unreliable models and erroneous scientific conclusions.
	
	The literature on robust variable selection has largely focused on casewise contamination (see, e.g., the overview in \citealp{maronna2019robust}). While these methods can provide protection against a limited number of atypical observations, cellwise contamination can propagate across variables and affect many observations, even when only a small fraction of cells are contaminated \citep{alqallaf2009propagation}. As a result, methods that are robust to casewise contamination may still perform poorly in the presence of cellwise outliers \citep{raymaekers2024challenges}. As in the casewise setting, it is useful to distinguish between outliers that are evident in individual variables and those that only emerge through multivariate relationships \citep{raymaekers2021handling}. The latter are particularly challenging because they distort the dependence structure of the data and can mislead variable selection methods that rely on it.
	
	Another important challenge arising in high-dimensional settings is that strong predictor dependence and model uncertainty often make it difficult to identify a single, well-defined sparse model. Ensemble methods have emerged as an alternative strategy that combines information across multiple competing models. Existing ensemble frameworks span: (i) randomization-based methods, such as Random Forest \citep{breiman2001random} and Random GLM \citep{song2013random}; (ii) gradient boosting algorithms, like XGBoost \citep{chen2016xgboost}; and (iii) deterministic, competitive frameworks, which partition the predictor space among multiple sub-models through a structured optimization process \citep{christidis2020split, christidis2025multi}. Despite the broad range of existing ensemble frameworks, robustness to cellwise contamination remains an important gap in the literature. Even recent advances in robust ensembles have focused exclusively on the casewise setting \citep{christidis2026robust}. The integration of cellwise-robust estimation within scalable ensemble frameworks remains largely unexplored.
	
	In this paper, we address this gap by introducing the Fast and Scalable Cellwise-Robust Ensemble (FSCRE) algorithm, a robust sparse ensemble framework for high-dimensional data affected by cellwise outliers. To maintain computational efficiency, FSCRE is not based on a complex global optimization methodology. Instead, the proposed method dynamically partitions predictors into disjoint sub-models following three main stages:
	
	\begin{enumerate}[label=(\roman*)]
		\item a cellwise-robust preprocessing stage, to identify outliers and estimate a robust correlation structure;
		\item a proposal stage, where a cellwise-robust correlation-based algorithm identifies candidate predictors for competing models; and
		\item an update stage, in which candidate predictors compete across sub-models using a robust prediction cross-validation criterion. 
	\end{enumerate}
	The proposed framework then fits $K$ disjoint sub-models with an MM-estimator to the preprocessed data and predicts new observations by averaging the predictions from the $K$ fitted sub-models. 
	
	We analyze the computational scalability of the proposed framework and discuss its transformation properties. Extensive simulation studies, including mixture scenarios combining casewise and cellwise contamination, and a bioinformatics data application are used to evaluate FSCRE's variable selection and prediction performance. An efficient software implementation is available on CRAN.	
	
	The remainder of this paper is organized as follows. Section~\ref{sec:background} establishes the data contamination framework and provides a review of the related literature. Section~\ref{sec:method} presents the details of the proposed FSCRE algorithm, and Section~\ref{sec:theory} analyzes its computational scalability and transformation properties. Section~\ref{sec:simulation} contains an extensive simulation study comparing FSCRE to state-of-the-art competitors. In Section~\ref{sec:bioinformatics}, we demonstrate the practical utility of our method on a bioinformatics data application. Finally, Section~\ref{sec:summary} concludes with a summary and discussion of future work.
	
	\section{Problem Setting and Background}\label{sec:background}
	
	Consider the linear regression model
	\begin{equation}\label{eq:linmodel}
		\by = \bX\bbet + \boldsymbol{\varepsilon},
	\end{equation}
	where $\by \in \mathbb{R}^n$ is the response vector, $\bX \in \mathbb{R}^{n \times p}$ is the predictor matrix, $\bbet \in \mathbb{R}^p$ is a sparse vector of unknown coefficients, and $\boldsymbol{\varepsilon} \in \mathbb{R}^n$ is a noise vector with i.i.d.\ entries. We focus on the high-dimensional regime $p \gg n$, where the objective is accurate variable selection and prediction under limited sample sizes.
	
	In practice, we observe a predictor matrix $\mathbf{X}$ that may contain cellwise contamination, with contaminated cells commonly assumed to arise independently throughout the design matrix \citep{alqallaf2009propagation}. Under this independent contamination model, even a small fraction of contaminated cells can affect a large proportion of observations (rows) through contamination propagation, causing procedures designed to handle only casewise contamination to break down \citep{alqallaf2009propagation,raymaekers2024challenges}. 
	
	Moreover, cellwise contamination may generate marginal outliers (detectable univariately) or correlation outliers that primarily distort multivariate dependence while remaining marginally plausible \citep{raymaekers2021handling}. In addition to contamination in the predictor matrix $\mathbf{X}$, atypical response values in $\mathbf{y}$ may also occur; in regression settings, these are naturally viewed as observation-level (casewise) contamination.
	
	A growing literature on cellwise-robust estimation has emerged in recent years, with early contributions focusing on robust multivariate location and scatter estimation under both casewise and cellwise contamination \citep{agostinelli2015robust}. Within this literature, the Detect Deviating Cells (DDC) algorithm \citep{rousseeuw2018detecting} has become a foundational method for identifying and imputing cellwise outliers by leveraging dependence among variables. Subsequent work has developed fast robust correlation estimators and related scalable tools for high-dimensional settings \citep{tarr2016robust, loh2018high, raymaekers2021fast, pacreau2023robust}.
	
	Cellwise contamination has also motivated robust regression procedures. In the lower-dimensional setting, prominent approaches include the Shooting S-estimator \citep{ollerer2016shooting} and Cellwise Robust M-regression (CRM) \citep{filzmoser2020cellwise}, the latter of which employs specialized diagnostics to identify contaminated cells \citep{debruyne2019outlyingness}. In the sparse high-dimensional setting, comparatively few methods have been proposed, with Sparse Shooting S \citep{bottmer2022sparse} and CR-Lasso \citep{su2024cr} among the main contributions. While these approaches represent important advances, they remain single-model estimators and do not leverage the stability and variance-reduction advantages of ensemble-based strategies.

	\section{The Fast and Scalable Cellwise-Robust Ensemble Algorithm}\label{sec:method}
	
	This section presents the proposed Fast and Scalable Cellwise-Robust Ensemble (FSCRE) algorithm, a framework for variable selection and prediction in high-dimensional settings under cellwise contamination. The central idea of FSCRE is to combine cellwise outlier detection and robust estimation with a competitive ensemble strategy while maintaining computational efficiency. As illustrated in Figure~\ref{fig:fscre_architecture} and summarized in Algorithm~\ref{alg:fscre_main}, the algorithm partitions the predictor space through a hybrid proposer--arbiter mechanism rather than constructing ensemble members independently. The procedure consists of three connected but independently structured stages. The following subsections describe these stages, followed by the ensemble prediction rule.
	
	\subsection{Cellwise-Robust Preprocessing and Correlations}\label{sec:method_foundation}
	
	The first stage of the FSCRE algorithm preprocesses the observed data matrix $[\by,\bX]$ to mitigate the effects of cellwise contamination before variable selection. The goal of this stage is to identify potentially contaminated cells, obtain an imputed data matrix, and estimate a robust correlation structure for subsequent stages of the procedure. 
	
	In the regression setup, the response is a univariate variable with either an outlying or a clean cell, while the predictor variables are jointly analyzed and cellwise outliers are jointly detected in the predictor matrix $\bX$ \citep{raymaekers2024challenges}. In our implementation, outliers in $\bX$ are detected and imputed using the Detect Deviating Cells (DDC) algorithm \citep{rousseeuw2018detecting} due to its strong empirical performance and scalability in high-dimensional settings, although the framework is not restricted to this particular choice. For scalability in high-dimensional settings, we use the fast implementation of DDC proposed by \citet{raymaekers2021fast} to reduce the computational complexity. This produces the first output of the preprocessing stage, an imputed matrix $\bX_{\text{imp}}$ to be used later for cross-validation and for the final regression fits.    
	
	We then estimate robust correlations using the transformation-based approach of \citet{raymaekers2021fast}: we apply the wrapping transformation to $\bX_{\text{imp}}$ and to the response vector to obtain wrapped data $(\bX_w,\by_w)$, and compute standard product-moment correlations on the wrapped data. As illustrated in Figure~\ref{fig:fscre_architecture}, this preprocessing stage yields a robust, positive semidefinite predictor correlation matrix $\mathbf{R}_{X}$ and a predictor--response correlation vector $\mathbf{r}_{y}$, which serve as the primary inputs for the selection engine.	
	
	\subsection{Identification of Candidate Predictors by Cellwise-Robust LARS}\label{sec:method_proposer}
	
	To identify predictors for each competing ensemble sub-model, we build on the correlation-based, casewise-robust Least-Angle Regression (LARS) algorithm of \citet{khan2007robust}. Specifically, the proposed procedure is formulated in terms of the cellwise-robust correlation quantities $(\mathbf{R}_{X}, \mathbf{r}_{y})$ developed in Section~\ref{sec:method_foundation}, thus providing protection against cellwise contamination. In addition, it operates entirely in the correlation space, avoiding repeated operations on the original $n \times p$ data matrix and thereby promoting computational scalability. 
	
	For each sub-model $k$, the procedure uses the global correlation matrix $\mathbf{R}_X$, the set of currently available predictors $V$, the current active set $S_k$, and its dynamic correlation vector $\mathbf{r}^{(k)}$. Restricting attention to the predictors in $V$, the next predictor $j_k^*$ to enter the current active set is identified along the robust LARS path by computing the minimum step size $\gamma_k^*$ in the equiangular direction. This selection procedure returns the quantities $(j_k^*, \gamma_k^*, \{a_j^{(k)}\})$, where $\{a_j^{(k)}\}$ denotes the correlation inner products used to update the correlation state in the next iteration. For the geometric derivations and algebraic details of the robust LARS path, we refer the reader to \citet{khan2007robust}. 
	
	As illustrated in Figure~\ref{fig:fscre_architecture}, the first phase (Phase~\ref{phase:propose}) of the procedure is performed in parallel to identify the best candidate predictor for each active set $S_k$. In Phase~\ref{phase:arbiter}, described in Section~\ref{sec:method_arbitration}, the arbiter algorithm determines which sub-model incorporates the selected predictor into its active set.
	
	\begin{figure}[h!]
		\centering
		\begin{tikzpicture}[
			font=\small,
			node distance=1.5cm and 2cm,
			box/.style={rectangle, draw, rounded corners, align=center, minimum height=0.8cm, fill=blue!5, thick},
			arbiter/.style={rectangle, draw, rounded corners, align=center, minimum height=1cm, fill=red!10, thick},
			state/.style={rectangle, draw, align=center, fill=gray!10, dashed},
			arrow/.style={->, >=Stealth, thick},
			looparrow/.style={->, >=Stealth, thick, dashed}
			]
			
			\node[box] (data) {\textbf{\textcolor{algblue}{Input}} \\[1ex] Observed Data \\$\bX, \by$};
			\node[box, below=0.8cm of data] (preproc) {\textbf{\textcolor{algblue}{Cellwise-Robust Foundation}} \\[1ex]
				$\begin{aligned}
					\bX &\xrightarrow{\text{DDC}} \bX_{\text{imp}} \\
					\left(\bX_{\text{imp}}, \by \right) &\xrightarrow{\text{Wrap}} \left(\mathbf{R}_X, \mathbf{r}_y \right)
				\end{aligned}$};
			
			\draw[arrow] (data) -- (preproc);
			
			\node[state, below=1.2cm of preproc] (init) {\textbf{\textcolor{algblue}{Initialize}} \\[1ex] Available Pool: $V \subseteq \{1, \dots, p\}$ \\[1ex] Sub-models $k \in \{1, \dots, K\}$: $S_k, \mathbf{r}^{(k)}$};
			
			\node[box, below=1cm of init, xshift=-3.2cm] (prop1) {\textbf{\textcolor{algblue}{Phase 1: Propose}} \\[1ex] Robust LARS 1 \\ $\to (j_1^*, \gamma_1^*)$};
			\node[box, below=1cm of init, xshift=3.2cm] (propK) {\textbf{\textcolor{algblue}{Phase 1: Propose}} \\[1ex] Robust LARS $K$ \\ $\to (j_K^*, \gamma_K^*)$};
			\node[below=1.5cm of init] (dots) {$\dots$};
			
			\node[arbiter, below=2.1cm of dots] (arbiter) {\textbf{\textcolor{algblue}{Phase 2: CV Arbiter}} \\[1ex]
				$B_k \to \text{CV-Error}(S_k)$ - $\text{CV-Error}(S_k \cup \{j_k^*\})$};
			
			\node[box, below=1.5cm of arbiter] (update) {\textbf{\textcolor{algblue}{Phase 3: Update}} \\[1ex] $k^* = \argmax_{k \in \{1,\dots,K\}} B_k$ \\[1ex] $S_{k^*} \gets S_{k^*} \cup \{j_{k^*}^*\}$ \\[1ex] $V \gets V \setminus \{j_{k^*}^*\}$ \\[1ex] Update $\mathbf{r}^{(k^*)}$};
			
			\coordinate (loopleft) at ([xshift=-0.6cm]prop1.west);
			\coordinate (loopright) at ([xshift=0.4cm]propK.east);
			
			\node[box, right=1.2cm of loopright |- arbiter] (final) {\textbf{\textcolor{algblue}{Phase 4: Final Fit}} \\[1ex] MM-estimators \\ on $S_1, \dots, S_K$};
			
			\begin{scope}[on background layer]
				\node[draw, dashed, thick, inner sep=0.6cm, fill=gray!5, 
				fit=(init) (prop1) (propK) (arbiter) (update) (loopleft) (loopright)] (loopbox) {};
				\node[above left, inner sep=0.2cm] at (loopbox.south east) {\textit{Iterative Selection Loop}};
			\end{scope}
			
			\draw[arrow] (preproc) -- (init);
			\draw[arrow] (init) -- (prop1);
			\draw[arrow] (init) -- (propK);
			\draw[arrow] (prop1) -- (arbiter);
			\draw[arrow] (propK) -- (arbiter);
			
			\draw[arrow] (arbiter) -- node[right, xshift=0.1cm] {$\max B_k > 0$} (update);
			
			\draw[looparrow] (update.west) -- (update.west -| loopleft) |- (init.west);
			
			\draw[arrow] (arbiter.east) -- node[above, pos=0.4] {$\max B_k \le 0$} (final.west);
			
		\end{tikzpicture}
		\caption{Schematic overview of the FSCRE architecture, integrating a cellwise-robust foundation with a competitive proposer--arbiter loop that partitions predictors into disjoint sub-models.}
		\label{fig:fscre_architecture}
	\end{figure}
	
	\subsection{Competitive Construction of Ensemble Sub-Models}\label{sec:method_arbitration}
	
	\begin{algorithm}[h!]
		\caption{The Fast and Scalable Cellwise-Robust Ensemble (FSCRE) Algorithm}
		\label{alg:fscre_main}
		\setcounter{algphase}{0} 
		\begin{algorithmic}[1]
			\Require Observed data $(\bX, \by)$; Number of models $K$; Tolerance $\tau$.
			\Statex
			
			\State \textbf{\textcolor{algblue}{--- Cellwise-Robust Foundation ---}}
			\Statex Obtain DDC-imputed $\bX_{\text{imp}}$ and wrapped $\by_w$; compute robust correlations $(\mathbf{R}_{X}, \mathbf{r}_{y})$.
			\Statex
			
			\State \textbf{\textcolor{algblue}{--- Initialize ---}} 
			\Statex \hspace{1em} Sub-models $k=1, \dots, K$: $S_k \gets \emptyset$, $\mathbf{s}_k \gets \emptyset$, and $\mathbf{r}^{(k)} \gets \mathbf{r}_{y}$.
			\Statex \hspace{1em} Available pool of predictors $V \gets \{1, \dots, p\}$.
			\Statex
			
			\While{$V \neq \emptyset$}
			\algphase{Propose}\label{phase:propose}
			\For{$k = 1, \dots, K$}
			\State Let $(j_k^*,\, \gamma^*_k,\, \{a_j^{(k)}\})$ be the proposal from the Robust LARS module for model $k$.
			\EndFor
			\Statex
			
			\algphase{CV Arbiter}\label{phase:arbiter}
			\For{$k = 1, \dots, K$}
			\If{$j_k^* \neq \text{null}$}
			\State $B_k \gets \text{CV-Error}(S_k) - \text{CV-Error}(S_k \cup \{j_k^*\})$ \Comment{Evaluate robust predictive benefit}
			\Else
			\State $B_k \gets -\infty$ \Comment{Model cannot advance further}
			\EndIf
			\EndFor
			\Statex
			
			\algphase{Update}\label{phase:update}
			\State $k^* \gets \argmax_{k \in \{1,\dots,K\}} B_k$ \Comment{Identify best overall sub-model move}
			\State Let $B^* \gets B_{k^*}$, $j^* \gets j_{k^*}^*$, $\gamma^* \gets \gamma_{k^*}^*$,  $\{a_j^*\} \gets \{a_j^{(k^*)}\}$
			\State $\mathcal{E}_{k^*} \gets \text{CV-Error}(S_{k^*})$
			\If{$B^* > 0$ \textbf{and} $B^* / \mathcal{E}_{k^*} > \tau$}
			\State \textbf{\textcolor{algblue}{Assign Winner and Update State:}}
			\State \hspace{1em} $S_{k^*} \gets S_{k^*} \cup \{j^*\}$ \Comment{Update winning model's active set}
			\State \hspace{1em} Add sign of predictor $j^*$ to $\mathbf{s}_{k^*}$
			\State \hspace{1em} $V \gets V \setminus \{j^*\}$ \Comment{Remove winner from available pool}
			\State \hspace{1em} $\mathbf{r}^{(k^*)}_j \gets \mathbf{r}^{(k^*)}_j - \gamma^* \cdot a_j^*$,\enspace for all $j \in V$ \Comment{Update correlation state}
			\Else
			\State \textbf{break} \Comment{No sufficient improvement; proceed to Final Fit}
			\EndIf
			\EndWhile
			\Statex
			
			\algphase{Final Fit}\label{phase:final}
			\State Fit casewise-robust MM-estimators on $S_1, \dots, S_K$ using the preprocessed data.
			\Statex
			
			\Ensure The disjoint predictor sets $S_1, \dots, S_K$ and final fitted models.
		\end{algorithmic}
	\end{algorithm}
	
	Having identified one candidate predictor for each sub-model, the second phase determines which active set $S_k$ is expanded in the current iteration (see Figure~\ref{fig:fscre_architecture}).
	
	For each proposal, $S_{k} \cup \{j_k^*\}, \; \text{for}\; k = 1, \ldots, K$, the algorithm computes its predictive benefit as the reduction in $v$-fold cross-validated prediction error relative to that of the model based on the current active set $S_k$, evaluated on the preprocessed training data $(\bX_{\text{imp}},\by_w)$. Since this scoring step is executed repeatedly within the inner loop, computational efficiency is essential. We therefore employ Huber M-estimation to fit models for each training set, which provides additional robustness to the scoring procedure while offering a practical compromise between robustness and computational cost. Fold errors are evaluated using the Huber loss $\rho_c(\cdot)$ \citep{maronna2019robust}.
	
	In the third phase (Phase~\ref{phase:update}), the proposal achieving the largest positive improvement is accepted, the corresponding predictor is assigned to the winning active set, and the predictor is removed from the pool of available predictors $V$ to enforce disjointness. The algorithm terminates when no proposal yields a positive improvement, or when the best relative improvement falls below the tolerance $\tau$.
	
	Although FSCRE is designed primarily to address cellwise contamination, additional casewise outliers may still be present. To account for this possibility, once the selection procedure partitions the predictors into disjoint sets $S_1,\dots,S_K$, the final phase (Phase~\ref{phase:final}) of the algorithm fits a casewise-robust MM-estimator \citep{yohai1987high} to the preprocessed data for each set. This final fitting step requires that $p_k = |S_k| < n$, which is inherently satisfied in sparse settings.
	
	\subsection{Ensemble Prediction}\label{sec:method_ensemble}
	
	Predictions for a new observation are obtained by averaging the predictions from the $K$ fitted sub-models. In the simulation study, predictive accuracy is evaluated on an independent uncontaminated test set.
	
	The number of sub-models $K$ controls the complexity of the ensemble. Larger values of $K$ tend to distribute correlated predictors across different disjoint sub-models, reducing the bias associated with a single greedy selection path while lowering prediction variance through aggregation \citep{ueda1996generalization}. In practice, values of $K \in \{5,\dots,10\}$ provide a favorable trade-off, as supported by the sensitivity analysis in Appendix~\ref{sec:sensitivity}.

	\section{Scalability and Transformation Properties}\label{sec:theory}
	
	This section analyzes the computational complexity of FSCRE and briefly discusses transformation properties of the framework. Together, these properties help characterize the behavior of FSCRE in high-dimensional applications.
	
	\subsection{Computational Complexity}\label{prop:complexity}
	
	Let $n$ be the number of observations, $p$ the number of predictors, $K$ the number of ensemble sub-models, $k_{\max}$ the total number of selected variables across the ensemble, and $\bar{s}$ the average sub-model size. In the typical $p \gg n$ sparse setting, the overall computational complexity of FSCRE is
	\[
	O\!\left( np^2 \;+\; k_{\max}\,K\left(p\bar{s} \;+\; v\,n\,\bar{s}^2\right)\right),
	\]
	where $v$ is the number of cross-validation folds.
	
	The first term corresponds to the construction of the robust correlation inputs. In our implementation, the predictor matrix is first cleaned using the fast DDC procedure \citep{rousseeuw2018detecting,raymaekers2021fast} and then wrapped to yield robust, positive semidefinite correlation estimates \citep{raymaekers2021fast}. Forming the full $p\times p$ correlation matrix $\mathbf{R}_X$ from the wrapped data requires $O(np^2)$ operations and $O(p^2)$ storage. In practice, this step reduces to a dense matrix multiplication and is highly optimized in modern numerical linear algebra libraries.
	
	The second term governs the iterative proposer--arbiter loop. At each of the $k_{\max}$ iterations, the robust LARS proposer is called once per sub-model at cost $O(p\bar{s})$, arising from correlation-space updates and searching over the available predictors. The arbiter evaluates $K$ candidate moves using $v$-fold cross-validation. For linear models fitted on $\bar{s}$ predictors, the resulting cost is $O(K\,v\,n\,\bar{s}^2)$ per iteration (up to constant factors, including the few IRLS iterations required to compute the Huber M-estimates during cross-validation). The final MM fits on the selected sets add $O(K\,n\,\bar{s}^2)$ operations, independent of $p$. A detailed derivation is provided in Appendix~\ref{sec:supp_complexity}, and empirical timing results are reported in Section~\ref{sec:simulation}.
	
	\subsection{Transformation Properties}
	
	Since FSCRE performs variable selection using correlation-based operations computed from the imputed data matrix, it naturally inherits the transformation properties of the underlying preprocessing procedure. In our current implementation, the preprocessing stage is based on DDC, which is equivariant under per-column affine transformations and predictor permutations \citep{rousseeuw2018detecting}. Similarly, the wrapping-based correlation estimates \citep{raymaekers2021fast} are invariant to per-variable shifts and rescalings, while guaranteeing positive semidefiniteness. As a result, variable-selection decisions are unaffected by per-variable shifts and rescalings, and transform accordingly under predictor permutations. Moreover, the correlation-based formulation of both the LARS proposer and the cross-validation criterion implies that variable selection is invariant to the inclusion of an intercept term in the internal regression models used by the arbiter.
	
	Finally, the sub-models are estimated using MM-regression, which is regression equivariant and affine equivariant with respect to predictor rescaling. Therefore, transformations of the predictors induce only the corresponding transformations of the fitted model coefficients, while leaving the resulting fitted values unchanged. Thus, the transformation properties established for the selection stage are preserved throughout the final model-fitting stage. Overall, these properties ensure that FSCRE behaves objectively under routine data transformations.
	
	\section{Simulation Study}\label{sec:simulation}
	
	To validate the performance of the FSCRE algorithm, we conduct a comprehensive simulation study designed to mimic challenging high-dimensional regression scenarios. The study evaluates variable selection accuracy and predictive performance against a suite of state-of-the-art competitors and baseline methods across a variety of contamination settings, ranging from classical casewise outliers to structured cellwise models that distort the correlation structure. In addition, we empirically evaluate computational time as both the sample size and the number of predictors increase.
	
	\subsection{Data Generation and Contamination Models}\label{sec:simulation_contamination}
	
	We generate data from the linear model $\by = \bX\bbet + \boldsymbol{\varepsilon}$ with $n=50$ and $p=500$. The rows of the clean design matrix $\bX$ are drawn from $\mathcal{N}_p(\mathbf{0}, \boldsymbol{\Sigma})$. To model complex collinearity, $\boldsymbol{\Sigma}$ has a block-diagonal structure: active predictors are grouped in blocks of size $25$ with correlation $\rho=0.8$, while the background correlation is $\rho_{0}=0.2$. The coefficient vector $\bbet$ is sparse, with $\lVert \bbet \rVert_0 \in \{50, 100, 200\}$ non-zero entries drawn uniformly from $[0, 5]$ with random signs. The error $\boldsymbol{\varepsilon}$ is scaled to achieve a target signal-to-noise ratio (SNR) in $\{0.5, 1, 2\}$. Predictive performance is evaluated on an independent uncontaminated test set. We examine the following contamination mechanisms, with $\mathcal{C}$ denoting the set of contaminated indices and $\alpha$ the contamination proportion. In the cellwise scenarios, contamination is applied to the predictor matrix $\bX$ only; response contamination is considered through casewise mechanisms.
	
	\begin{enumerate}
		\item \textbf{Casewise:} A proportion $\alpha \in \{0.1, 0.2\}$ of rows $i \in \mathcal{C}$ are replaced by high-leverage outliers $\bx_i \sim \mathcal{N}_p(\mathbf{0}, 0.1\mathbf{I}) + c \bu$, where $\bu$ is the smallest eigenvector of $\boldsymbol{\Sigma}$ and $c=2$. The response is set to $y_i = \bx_i^\top \bbet_{\text{cont}}$, where $\bbet_{\text{cont}}$ distorts active coefficients by a factor of $100$.
		
		\item \textbf{Cellwise Marginal:} A random subset of predictor cells $(i,j)\in \mathcal{C}$ with proportion $\alpha \in \{0.05, 0.1\}$ is replaced by $x_{ij} \sim \mathcal{N}(\mu', 1)$ with $\mu' = 10$, representing a large shift from the clean mean.
		
		\item \textbf{Cellwise Correlation:} This structured scenario targets the covariance structure. For a contaminated row $i$, a subset of predictor cells $J$ (overall proportion $\alpha \in \{0.05, 0.1\}$) is selected. Let $\boldsymbol{\Sigma}_{J}$ be the submatrix corresponding to these cells and let $\bv_{\min}$ be an eigenvector associated with the smallest eigenvalue of $\boldsymbol{\Sigma}_J$. The contaminated cells are replaced by a scaled version of $\bv_{\min}$ chosen to have a controlled Mahalanobis length with respect to $\boldsymbol{\Sigma}_J$, that is,
		\[
		\bx_{i,J} \;=\; \gamma \sqrt{|J|}\,
		\frac{\bv_{\min}}{\sqrt{\bv_{\min}^{\top}\boldsymbol{\Sigma}_J^{-1}\bv_{\min}}},
		\qquad \gamma=3.
		\]
		This aligns contamination with a low-variation direction and can remain marginally plausible while distorting dependence, making it difficult to detect with univariate filters.
		
		\item \textbf{Mixture Marginal:} A fraction $\alpha_1 = 0.1$ of rows are casewise contaminated as above, and the remaining rows are subject to cellwise marginal contamination at rate $\alpha_2 = 0.05$.
		
		\item \textbf{Mixture Correlation:} A fraction $\alpha_1 = 0.1$ of rows are casewise contaminated as above, and the remaining rows are subject to cellwise correlation contamination at rate $\alpha_2 = 0.05$.
	\end{enumerate}
	
	\subsection{Methods}\label{sec:simulation_methods}
	
	We compare the performance of FSCRE against a comprehensive suite of baseline and state-of-the-art methods. To isolate the effects of data cleaning, ensemble architecture, and robust estimation, the competitors are organized as follows:
	
	\begin{itemize}
		\item \textbf{Proposed Estimators:} We evaluate the full \textbf{FSCRE} algorithm ($K=10$) alongside a single-model variant, \textbf{Cellwise-Robust LARS (CellRLARS)} ($K=1$), implemented in our \texttt{srlars} package available on CRAN \citep{srlars_package}. This comparison isolates the specific performance gains attributable to the competitive partitioning architecture.
		
		\item \textbf{DDC-Augmented Baselines:} To contrast our integrated architecture with standard two-step sequential pipelines, we include \textbf{DDC-EN} (standard Elastic Net applied to DDC-imputed data) and \textbf{DDC-RGLM} (Random GLM applied to imputed data). We use the \texttt{cellWise} package \citep{cellWise_package} for DDC, \texttt{glmnet} \citep{glmnet_package} for the elastic net, and \texttt{randomGLM} \citep{randomGLM_package} for the ensemble. The latter serves as a strong benchmark, representing a generic cellwise-robustified ensemble.
		
		\item \textbf{Sparse and Cellwise-Robust Estimators:} We compare against leading purpose-built methods using their authors' implementations: the iterative \textbf{Sparse Shooting S (Sparse-S)} estimator \citep{bottmer2022sparse} and the regularization-based \textbf{CR-Lasso} \citep{su2024cr}.
		
		\item \textbf{Non-Robust Baseline:} The standard \textbf{Elastic Net (EN)}, implemented in \texttt{glmnet}, is included to quantify the impact of contamination on non-robust methods.
	\end{itemize}
	
	All methods utilizing DDC employ the scalable \texttt{fastDDC} option implemented in the \texttt{cellWise} package. Tuning parameters are selected using the default cross-validation procedures provided by the corresponding software implementations.
	
	For scenarios involving pure casewise contamination, we additionally include sparse least-trimmed squares (\textbf{SparseLTS}) \citep{alfons2013sparse}, implemented in \texttt{robustHD} \citep{robustHD_package}, and the adaptive penalized elastic net S-estimator (\textbf{PENSE}) \citep{kepplinger2023robust}, implemented in \texttt{pense} \citep{pense_package}. These methods serve as gold-standard benchmarks for casewise outliers; however, they are excluded from the cellwise and mixture scenarios due to their inherent lack of theoretical robustness to cellwise contamination and their prohibitive computational cost in such settings.
	
	\subsection{Performance Measures}\label{sec:simulation_metrics}
	
	We evaluate estimator performance using four metrics, averaged over $N=50$ simulation runs.
	
	\begin{enumerate}
		\item \textbf{Mean Squared Prediction Error (MSPE):} Computed on an independent, uncontaminated test set of size $m=5{,}000$, and scaled by the noise variance $\sigma^2$. For test response vector $\by_{\text{test}}$ and prediction $\hat{\by}_{\text{test}}$, we report
		\[
		\text{MSPE}
		\;=\;
		\frac{1}{\sigma^2}\cdot \frac{1}{m}\,\bigl\|\hat{\by}_{\text{test}}-\by_{\text{test}}\bigr\|_2^2.
		\]
		Under the correctly specified linear model, the lower bound of this scaled MSPE is $1$.
		
		\item \textbf{Recall (RC):} The proportion of true active coefficients correctly identified as non-zero. Let $\mathcal{A} = \{j : \beta_j \neq 0\}$ be the true active set and $\hat{\mathcal{A}} = \{j : \hat{\beta}_j \neq 0\}$ be the selected set. Then $\text{Recall} = |\hat{\mathcal{A}} \cap \mathcal{A}| / |\mathcal{A}|$.
		
		\item \textbf{Precision (PR):} The proportion of selected coefficients that are truly active, defined as $\text{Precision} = |\hat{\mathcal{A}} \cap \mathcal{A}| / |\hat{\mathcal{A}}|$. High precision indicates effective control of false discoveries.
		
		\item \textbf{Computation Time (CPU):} The average time in seconds to fit the model, serving as empirical validation of scalability.
	\end{enumerate}
	
	Note that variable selection metrics (RC and PR) are not reported for DDC-RGLM; because this specific baseline uses an ensemble of 100 randomly seeded models, it trivially selects nearly all variables across the ensemble, rendering selection metrics uninformative.

	\subsection{Results}
	
	The simulation grid encompasses a wide array of configurations across various contamination proportions, sparsity levels, and signal-to-noise ratios. To synthesize these results, we compute average relative ranks, which provide a concise summary of performance across settings.
	
	As detailed in Table~\ref{tab:sim_cellwise} for the cellwise and mixture scenarios, the proposed FSCRE framework consistently attains the best average rank (approximately $1.0$--$1.2$) in predictive accuracy (MSPE), while maintaining a competitive balance of recall and precision. These results suggest that FSCRE is well-suited to settings with strong block collinearity and structured contamination, where combining information across disjoint sub-models can help capture complementary signal components and yield stable out-of-sample performance. In comparison, approaches based on a single global optimization may be more sensitive to these data characteristics, which can translate into less stable predictive accuracy.
	
	\begin{table}[h!]
		\centering
		\caption{Average relative ranks of the evaluated methods across all cellwise and mixture contamination scenarios. A rank of $1$ indicates the best performance among the seven methods. The top two performing methods in each column are highlighted in bold.}
		\label{tab:sim_cellwise}
		\resizebox{\textwidth}{!}{
			\begin{tabular}{l ccc ccc ccc ccc}
				\toprule
				& \multicolumn{3}{c}{\textbf{Cellwise Marginal}} & \multicolumn{3}{c}{\textbf{Cellwise Correlation}} & \multicolumn{3}{c}{\textbf{Mixture Marginal}} & \multicolumn{3}{c}{\textbf{Mixture Correlation}} \\
				\cmidrule(lr){1-1} \cmidrule(lr){2-4} \cmidrule(lr){5-7} \cmidrule(lr){8-10} \cmidrule(lr){11-13}
				\textbf{Method} & MSPE & RC & PR & MSPE & RC & PR & MSPE & RC & PR & MSPE & RC & PR \\
				\cmidrule(lr){1-1} \cmidrule(lr){2-4} \cmidrule(lr){5-7} \cmidrule(lr){8-10} \cmidrule(lr){11-13}
				EN          & 4.6 & 5.0 & 5.4 & 3.6 & 3.1 & 4.5 & 4.8 & 5.8 & 5.9 & 4.8 & 5.6 & 4.8 \\
				\addlinespace
				DDC-EN      & 3.1 & 3.1 & 4.9 & 3.7 & 3.9 & 5.5 & \textbf{2.0} & 3.9 & 4.3 & \textbf{2.0} & 3.9 & 4.6 \\
				DDC-RGLM    & \textbf{1.8} & --  & --  & \textbf{1.8} & --  & --  & 3.8 & --  & --  & 3.6 & --  & --  \\
				\addlinespace
				Sparse-S    & 7.0 & \textbf{1.7} & \textbf{1.1} & 6.8 & \textbf{1.5} & \textbf{1.0} & 6.7 & \textbf{1.8} & \textbf{1.0} & 6.4 & \textbf{1.4} & \textbf{1.0} \\
				CR-Lasso    & 6.0 & 4.0 & 4.6 & 6.2 & 5.0 & 4.9 & 6.3 & 3.1 & 4.8 & 6.6 & 3.1 & 5.7 \\
				\addlinespace
				CellRLARS       & 4.3 & 5.9 & \textbf{1.9} & 4.7 & 6.0 & \textbf{2.0} & 3.4 & 5.2 & \textbf{2.0} & 3.7 & 5.4 & \textbf{2.0} \\
				\textbf{FSCRE} & \textbf{1.2} & \textbf{1.3} & 3.0 & \textbf{1.2} & \textbf{1.5} & 3.1 & \textbf{1.0} & \textbf{1.2} & 3.0 & \textbf{1.0} & \textbf{1.6} & 3.0 \\
				\bottomrule
			\end{tabular}
		}
	\end{table}
	
\begin{figure}[h!]
		\centering
		\includegraphics[width=0.9\textwidth]{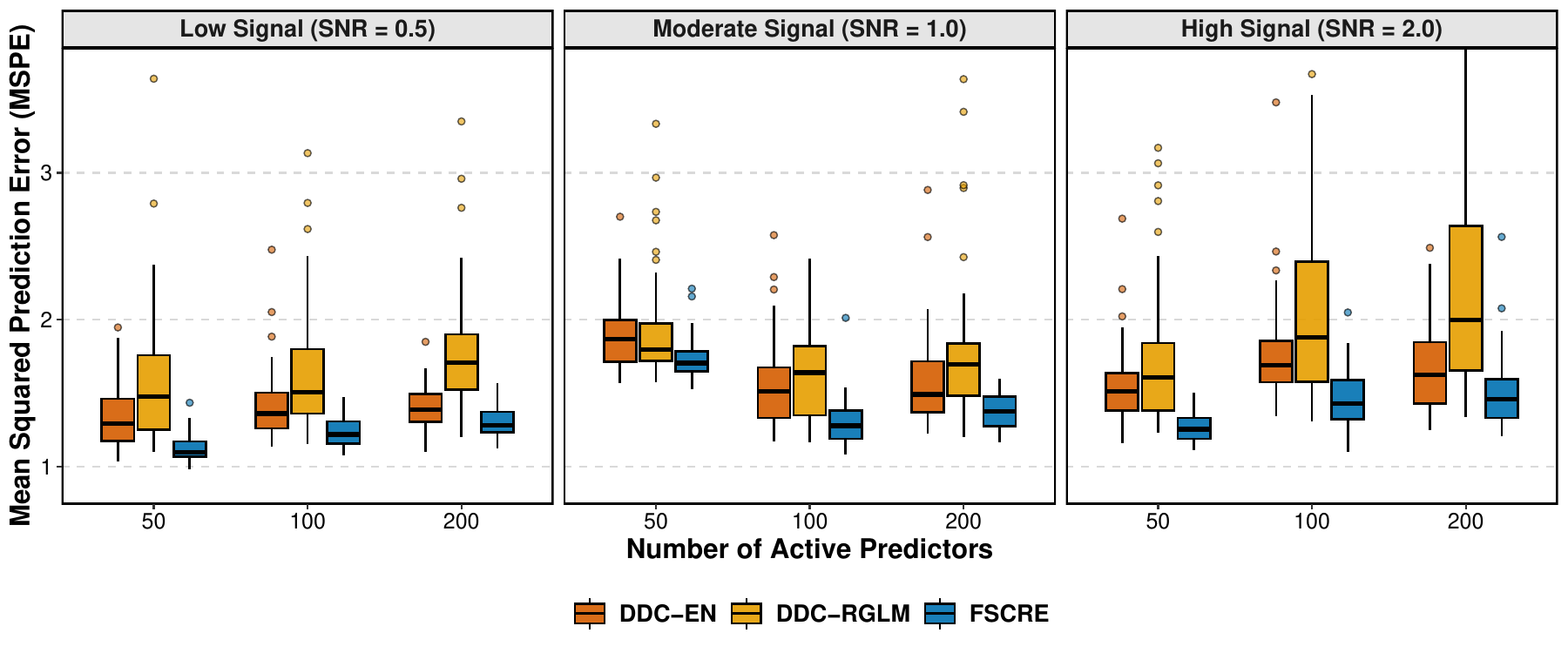}
		\caption{MSPE across 50 splits for the \textit{Mixture Correlation} scenario. Performance is evaluated across three SNRs and three sparsity levels. Because the error is scaled by the noise variance, the optimal possible MSPE is 1.0.}
		\label{fig:sim_mspe}
	\end{figure}

\begin{figure}[h!]
		\centering
		\includegraphics[width=0.9\textwidth]{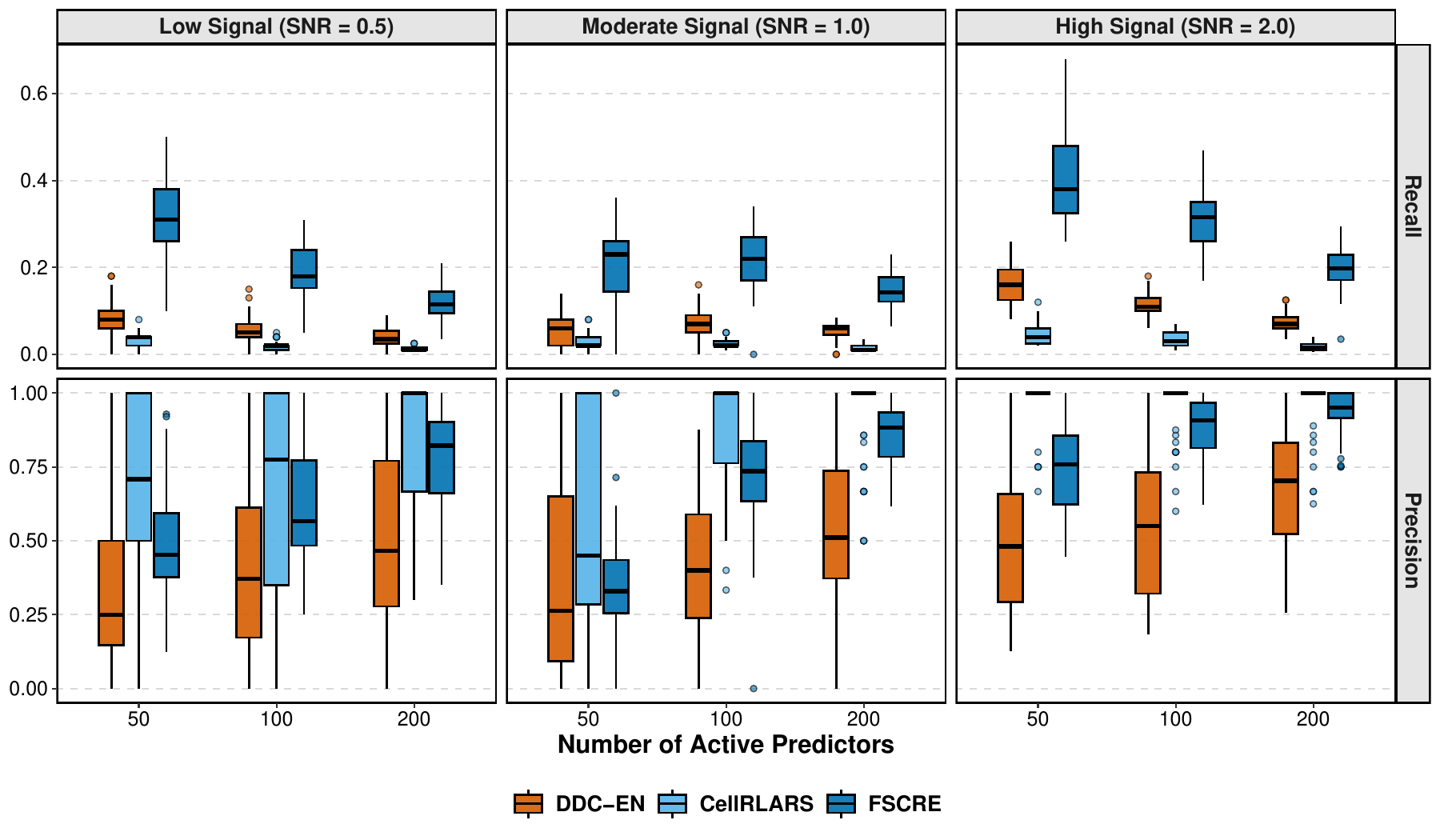}
		\caption{Recall (top row) and precision (bottom row) across 50 splits for the \textit{Mixture Correlation} scenario.}
		\label{fig:sim_rcpr}
	\end{figure}

	To visualize absolute performance, Figure~\ref{fig:sim_mspe} shows the MSPE distributions for the leading methods under a representative \textit{Mixture Correlation} scenario. Because MSPE is scaled by the noise variance, the theoretical lower bound is $1.0$. Across SNRs and sparsity levels, FSCRE attains the lowest median prediction error and is typically closest to this bound using only $K=10$ deterministic sub-models. This predictive performance is consistent with its variable selection behavior (Figure~\ref{fig:sim_rcpr}). In particular, relative to the single-model cellwise-robust baseline (CellRLARS), FSCRE achieves markedly higher recall and more stable precision, suggesting that relying on a single greedy path can be unreliable in the presence of dense collinearity and contamination. Moreover, the competitive allocation of predictors across disjoint sub-models appears to temper false discoveries while retaining relevant variables, leading to an improved balance of precision and recall compared with standard penalized baselines such as DDC-EN.
	
	Finally, to assess whether robustness to cellwise outliers compromises baseline performance, we evaluate all methods on uncontaminated data and under pure casewise contamination (Table~\ref{tab:sim_casewise}). On clean data, FSCRE achieves the best average MSPE rank, suggesting that the multi-model selection strategy can also be beneficial in the absence of contamination. Under casewise contamination, we compare FSCRE to PENSE and SparseLTS, which are widely used benchmarks for row-wise outliers. FSCRE remains competitive, with a better average MSPE rank than PENSE ($1.2$ versus $1.9$) and comparable recall and precision. Overall, these results indicate that FSCRE improves robustness to cellwise contamination while maintaining strong performance in clean and casewise-contaminated settings.
	
	\begin{table}[h!]
		\centering
		\caption{Average relative ranks of the evaluated methods for the uncontaminated (Clean) and pure Casewise contamination scenarios. A rank of $1$ indicates the best performance among the nine methods. The top two performing methods in each column are highlighted in bold.}
		\label{tab:sim_casewise}
		\begin{tabular}{l ccc ccc}
			\toprule
			& \multicolumn{3}{c}{\textbf{Clean Data}} & \multicolumn{3}{c}{\textbf{Casewise}} \\
			\cmidrule(lr){1-1} \cmidrule(lr){2-4} \cmidrule(lr){5-7}
			\textbf{Method} & MSPE & RC & PR & MSPE & RC & PR \\
			\cmidrule(lr){1-1} \cmidrule(lr){2-4} \cmidrule(lr){5-7}
			EN          & 5.3 & 6.1 & 4.8 & 6.4 & 7.5 & 5.4 \\
			\addlinespace
			DDC-EN      & 3.7 & 4.3 & 6.0 & 3.0 & 5.0 & 5.2 \\
			DDC-RGLM    & \textbf{2.6} & --  & --  & 5.0 & --  & --  \\
			\addlinespace
			Sparse-S    & 8.6 & \textbf{2.2} & \textbf{1.0} & 8.4 & \textbf{2.2} & \textbf{1.0} \\
			CR-Lasso    & 8.4 & 5.7 & 6.0 & 8.6 & 4.4 & 6.7 \\
			\addlinespace
			SparseLTS   & 6.8 & 6.0 & 8.0 & 4.7 & 5.6 & 7.9 \\
			PENSE       & 3.2 & \textbf{1.4} & 4.8 & \textbf{1.9} & \textbf{1.3} & 4.8 \\
			\addlinespace
			CellRLARS       & 5.4 & 7.9 & \textbf{2.0} & 5.7 & 7.5 & \textbf{2.1} \\
			\textbf{FSCRE} & \textbf{1.0} & {2.3} & 3.4 & \textbf{1.2} & 2.6 & 2.9 \\
			\bottomrule
		\end{tabular}
	\end{table}
	
	\subsection{Computational Scalability Study}\label{sec:simulation_scalability}
	
	To empirically assess the complexity bounds established in Section~\ref{sec:theory}, we conduct a computational scalability study. We consider the \textit{Mixture Correlation} setting ($\text{SNR}=1$, $\lVert \bbet_0 \rVert_0 = 50$), which is representative of a challenging regime with structured correlation outliers. Total CPU time is measured over a full-factorial grid, with sample size $n \in \{50,100,200,500\}$ and number of predictors $p \in \{50,100,500,1{,}000,2{,}000,5{,}000\}$.
	
\begin{figure}[h!]
		\centering
		\includegraphics[width=0.9\textwidth]{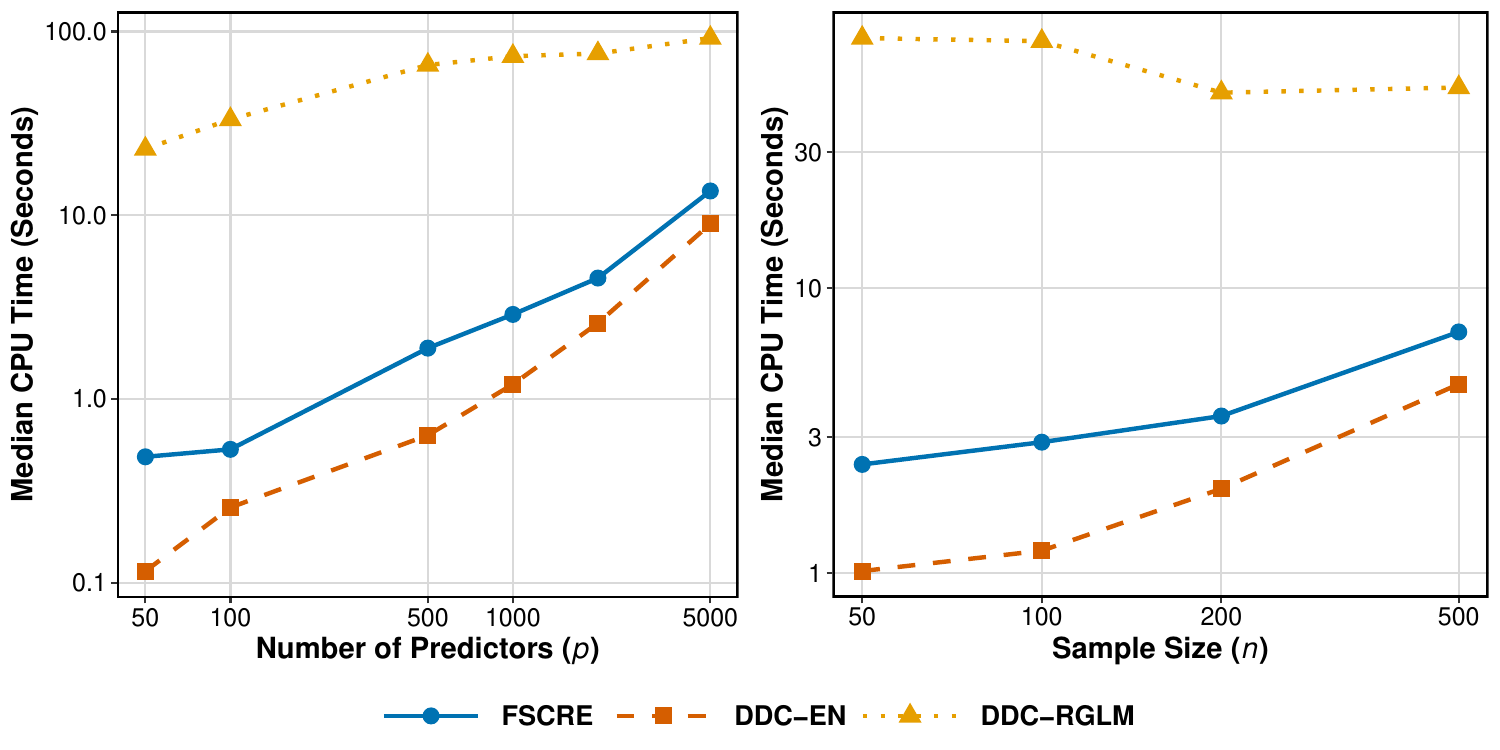}
		\caption{Median computational execution time (in seconds) of FSCRE and DDC-EN, plotted on logarithmic axes. \textbf{Left:} Time as a function of the number of predictors $p$, with sample size fixed at $n=100$. \textbf{Right:} Time as a function of sample size $n$, with predictors fixed at $p=1{,}000$.}
		\label{fig:cpu_scaling}
	\end{figure}

	As benchmarks, we compare the full FSCRE ensemble ($K=10$) to a sequential baseline (DDC-EN) and a generic robustified ensemble (DDC-RGLM). Each of the 24 configurations is replicated 50 times. The observed execution times are consistent with the scalability analysis in Section~\ref{sec:theory}. Figure~\ref{fig:cpu_scaling} displays two representative slices on logarithmic axes.
	
	In particular, the left panel indicates that FSCRE and DDC-EN exhibit the expected near-linear scaling with $p$. Across the range of problem sizes, FSCRE tracks closely with DDC-EN, with a relatively modest and approximately constant multiplicative overhead. For example, with $n=100$ and $p=5{,}000$, FSCRE completes in just over 10 seconds on average, whereas DDC-RGLM requires close to 100 seconds. Overall, these results suggest that the structured FSCRE architecture (correlation-space proposing, robust CV-based arbitration, and final MM fitting) can deliver ensemble-level robustness while remaining computationally comparable to a single penalized regression fit on imputed data.
	
	\section{Bioinformatics Data Application}\label{sec:bioinformatics}
	
	To demonstrate the practical utility and robustness of the FSCRE algorithm on real-world high-dimensional data, we apply it to a proteogenomics prediction task. Although the Central Dogma of Biology states that mRNA is translated into protein, the correlation between mRNA expression and protein abundance is often low to modest \citep{vogel2012insights, fortelny2017predict}. This discrepancy is driven not only by post-translational regulation but also by technical artifacts in genomic and proteomic measurements (e.g., missingness, varying detection limits, and intensity spikes) \citep{karpievitch2012normalization}, which closely mirror the cellwise contamination paradigm.
	
	\subsection{Data Description and Preprocessing}
	
	We utilize matched transcriptomic and proteomic data from the Breast Invasive Carcinoma (BRCA) cohort of The Cancer Genome Atlas (TCGA; \citealp{weinstein2013cancer}), accessed via the \texttt{curatedTCGAData} package \citep{curatedTCGAData_package}. Our objective is to predict the protein abundance of a key cancer driver using the global mRNA expression profile. We select the Estrogen Receptor alpha (ER-$\alpha$) as our target protein. ER-$\alpha$ is the defining biomarker and primary therapeutic target for the luminal subtypes, which represent the vast majority of breast cancer cases \citep{prat2015clinical}, and its expression is known to have a strong transcriptional basis, providing a reliable biological signal to model.
	
	The response vector $\by$ consists of standardized Reverse Phase Protein Array (RPPA) expression values for ER-$\alpha$, and the predictor matrix $\bX$ consists of the corresponding RNA-sequencing measurements. After matching samples with both RNA and protein data and removing observations with missing response values, the final dataset comprises $n=882$ observations.
	
	For each train/test split described in Section~\ref{sec:bioinformatics_contam}, we remove genes with near-zero variance using the training data only, retain the $p=500$ genes with the highest absolute marginal correlation with the target protein (computed on the training set only), and standardize predictors and response using training-set means and standard deviations. The resulting transformations are then applied to the corresponding test set.
	
	\subsection{Targeted Contamination Strategy}\label{sec:bioinformatics_contam}
	
	We evaluate predictive performance against the competitor methods (Section~\ref{sec:simulation_methods}) across $50$ random train/test splits. Each split allocates $n=50$ observations to training and the remaining $m \approx 830$ to testing, thereby enforcing a challenging high-dimensional regime ($p=500 \gg n=50$). The models are evaluated under two conditions:
	\begin{enumerate}
		\item \textit{Original data:} Models are trained on the unadulterated splits to establish baseline performance under natural biological and technical variability.
		\item \textit{Targeted artificial contamination:} In highly collinear data, $\ell_1$-penalized methods can often bypass random outliers by selecting correlated proxies. To evaluate robustness when the primary signal-carrying features are corrupted, we introduce contamination targeted to predictive genes. For each split, we fit a standard Elastic Net model on the training set and identify the $30$ most predictive genes. We then inject cellwise outliers ($\pm 10$ standard deviations) into $15\%$ of the training-set cells within these columns.
	\end{enumerate}
	Under this scenario, a robust method should retain predictive stability and recover the underlying biological signal rather than relying on proxy features.
	
	\subsection{Results and Biological Interpretability}\label{sec:bioinformatics_results}
	
\begin{figure}[h!]
		\centering
		\includegraphics[width=0.675\textwidth]{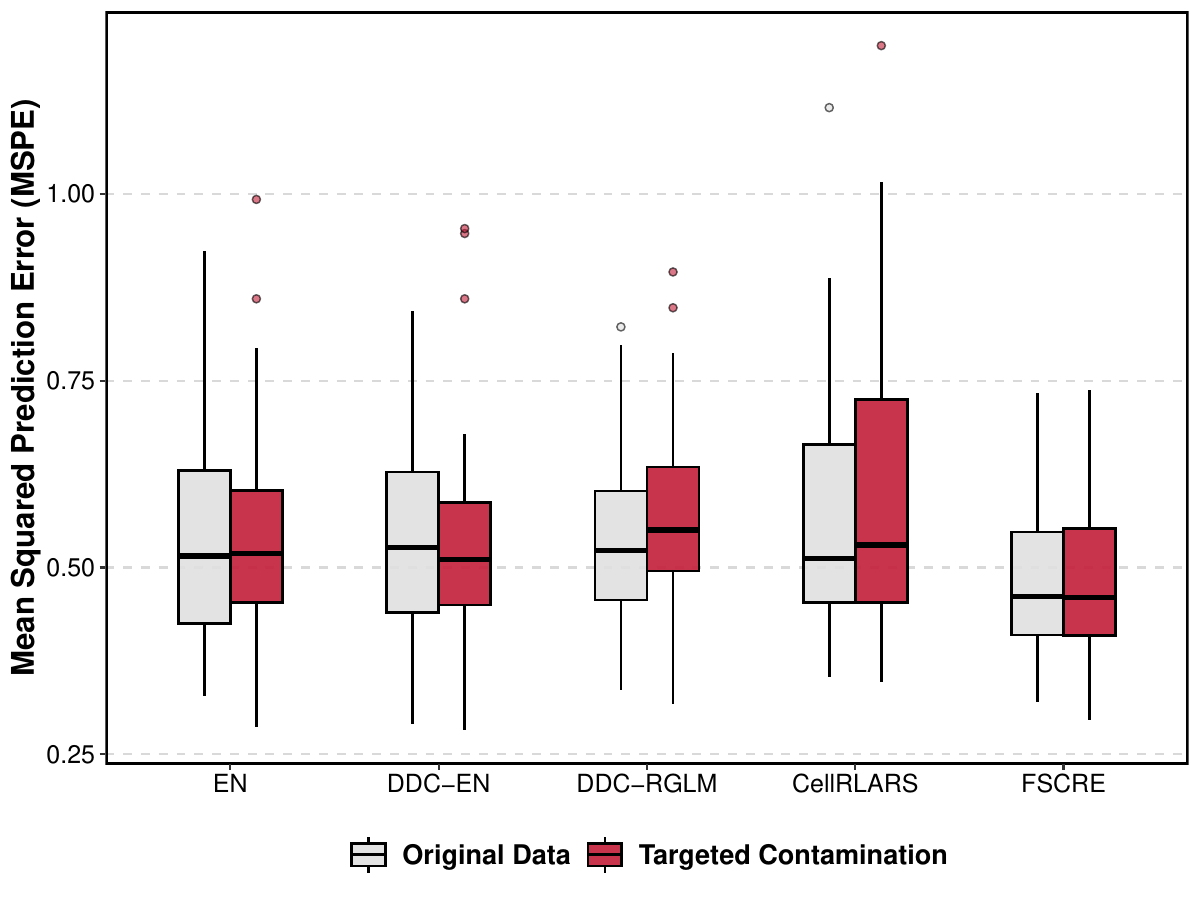}
		\caption{MSPE across 50 random splits for the prediction of ER-$\alpha$ protein abundance. The models are evaluated on both the original TCGA data (light grey) and data subjected to targeted artificial cellwise contamination in the training set (dark red).}
		\label{fig:app_mspe}
	\end{figure}

	Predictive performance across the 50 random splits for the ER-$\alpha$ target is shown in Figure~\ref{fig:app_mspe}. The single-model robust estimators (Sparse-S and CR-Lasso) are omitted from the figure; consistent with the simulation study, they frequently encounter convergence difficulties under the strong collinearity of the real data, leading to unstable predictions with MSPE values well outside the plotted range. On the original data, all methods indicate a substantial predictive signal, with FSCRE attaining the lowest median MSPE.
	
	The targeted contamination experiment provides a more stringent robustness assessment. Under this setting, the MSPE distributions of Elastic Net and the DDC-based baselines change relatively little. In highly collinear genomic data, this behavior is consistent with penalized methods shifting from corrupted primary predictors to correlated proxies, which can preserve predictive accuracy while reducing interpretability. The single-model robust baseline (RLARS) exhibits increased variability under contamination, suggesting sensitivity to the perturbed features. By comparison, FSCRE maintains a similar MSPE distribution across the two conditions and achieves the lowest median MSPE with comparatively low dispersion across splits.
	
	Beyond predictive accuracy, we examine the stability of selecting biologically meaningful variables under contamination. The target protein, ER-$\alpha$, is encoded by the \textit{ESR1} gene \citep{patel2022estrogen}. On the original data, Elastic Net selects \textit{ESR1} in 52\% of the 50 splits, dropping to 38\% under targeted contamination; DDC-EN exhibits a similar decline (54\% to 32\%). In contrast, FSCRE selects \textit{ESR1} in 90\% of splits on the original data and 84\% under targeted contamination, indicating substantially improved stability under cellwise corruption.
	
	We also consider secondary signals that may be masked in single-model fits. Under targeted contamination, FSCRE selects \textit{PADI2} in 42\% of splits (38\% on the original data); \textit{PADI2} has been implicated in ER-$\alpha$ transcriptional regulation \citep{zhang2012pad2}. FSCRE also selects \textit{PSAT1} in 42\% of splits in both conditions; \textit{PSAT1} has been associated with tamoxifen outcome in recurrent breast cancer \citep{marchi2017psat1}. Finally, FSCRE recovers established ER-positive/luminal signals such as \textit{TBC1D9} and the ER co-regulatory transcription factor \textit{GATA3} in $\approx 30\%$ of contaminated splits \citep{kothari2021tbc1d9, eeckhoute2007positive}, whereas Elastic Net and DDC-EN select \textit{PADI2}/\textit{PSAT1} in at most 8--10\% of contaminated splits and select \textit{TBC1D9}/\textit{GATA3} in $\leq 2\%$ of splits. Overall, these results suggest that the disjoint multi-model structure of FSCRE improves the stability of selecting biologically interpretable features under targeted cellwise contamination.

	\section{Summary and Future Work}\label{sec:summary}
	
	In this paper, we introduce the Fast and Scalable Cellwise-Robust Ensemble (FSCRE) algorithm to address a methodological gap in high-dimensional data analysis. While ensemble frameworks can improve stability under collinearity and model uncertainty, they typically lack resilience to cellwise contamination. FSCRE bridges this divide by combining cellwise outlier detection techniques, robust positive semidefinite correlation inputs, and a competitive proposer--arbiter architecture for constructing disjoint sparse sub-models.
	
	Extensive simulations demonstrate the benefits of this approach across a wide spectrum of contamination models, from classical casewise outliers to structured correlation outliers. By forcing predictors to compete across disjoint sub-models, FSCRE can regularize against false discoveries and yield favorable balances of precision and recall that translate into competitive predictive accuracy.
	
	These empirical findings are corroborated by our application to TCGA proteogenomics data. Under targeted artificial contamination designed to corrupt signal-carrying features, FSCRE maintains stable predictions. Moreover, the disjoint ensemble structure facilitates the recovery of secondary biological signals that were often masked in single-model alternatives.
	
	We also analyze the computational scalability and transformation properties of the framework. By structuring selection as a sequence of locally optimal competitive updates and operating in correlation space, FSCRE remains computationally efficient in high-dimensional settings.
	
	Several promising avenues remain for future research. First, the framework's modularity allows the integration of future advances in cellwise-robust preprocessing and correlation estimation. Second, relaxing the strict disjoint constraint through controlled variable sharing could capture more complex dependence structures. Finally, extending this architecture to generalized linear models, such as logistic or Cox regression, would broaden its applicability to high-dimensional classification and survival analysis tasks.
	
	\section*{Data and Code Availability}
	
	For full reproducibility, the scripts for the synthetic experiments, simulation study, and TCGA bioinformatics application are publicly available at \url{https://github.com/AnthonyChristidis/FSCRE-Simulations}. The proposed methodology is implemented in the \texttt{srlars} \texttt{R} package, available on CRAN.
	
	\section*{Conflict of Interest}
	
	The authors declare no potential conflicts of interest.

	\appendix
	
	\titleformat{\section}{\Large\bfseries}{Appendix \thesection:}{0.5em}{}
	
	\section{Derivation of Computational Complexity}\label{sec:supp_complexity}
	
	This section provides a step-by-step derivation of the computational complexity bound presented in Section~\ref{prop:complexity}.
	
	Let $n$ be the number of observations, $p$ the number of predictors, $K$ the number of ensemble models, $k_{\max}$ the total number of selected variables, and $\bar{s} \approx \frac{1}{K}\sum_{k=1}^K |S_k|$ the average sub-model size. In the typical $p \gg n$ sparse setting, the overall computational complexity of FSCRE is
	\[
	O\!\left( np^2 \;+\; k_{\max}\,K\left(p\bar{s} \;+\; v\,n\,\bar{s}^2\right)\right),
	\]
	where $v$ is the number of cross-validation folds. This bound is obtained by summing the costs incurred in the following stages: (1) preprocessing and correlation construction, (2) competitive selection, and (3) final model fitting.
	
	\textbf{1. Preprocessing and correlation construction:} The first stage consists of cellwise cleaning of the predictor matrix followed by robust correlation estimation. We use the fast DDC methodology of \citet{rousseeuw2018detecting,raymaekers2021fast} to identify and impute deviating predictor cells. In high-dimensional settings, the computational cost of fast DDC scales approximately as $O(np\log p)$. Next, robust correlations are computed using the transformation-based approach of \citet{raymaekers2021fast}: after applying the wrapping transformation to the preprocessed data, the predictor correlation matrix $\mathbf{R}_X$ is obtained by standard product-moment correlations on the wrapped predictors. Forming the full dense $p\times p$ correlation matrix requires $O(np^2)$ operations (and $O(p^2)$ storage). Consequently, the preprocessing phase is dominated by the correlation construction and has overall cost $O(np^2)$.
	
	\textbf{2. Competitive ensemble selection:} The iterative selection loop executes for at most $k_{\max}$ iterations. Within each iteration, each sub-model proposes a candidate predictor. For a model with active set size $s_k$, the proposer computes geometric quantities based on the signed correlation submatrix of dimension $s_k\times s_k$, which requires solving a linear system at cost $O(s_k^3)$ and subsequent updates at cost $O(s_k^2)$. The dominant cost arises from evaluating the correlation-based update quantities over the pool of available predictors $V$ with $|V|\approx p$, which costs $O(p s_k)$. In the sparse $p\gg n$ regime with $s_k \ll n$, the $O(p s_k)$ term dominates $O(s_k^3)$, yielding a proposal cost of $O(p s_k)$ per model. Summing over the $K$ models gives a total proposal cost per iteration of $O(K\,p\,\bar{s})$.
	
	The arbiter evaluates the $K$ proposed moves via $v$-fold cross-validation on the preprocessed data. Fitting a linear model on approximately $n$ observations with $s_k$ predictors costs $O(n s_k^2)$ (up to constant factors, including any small number of IRLS iterations when robust fitting is used in the arbiter). Therefore, the arbitration cost per iteration is $O(K\,v\,n\,\bar{s}^2)$. Multiplying by $k_{\max}$ yields the total selection-loop cost
	\[
	O\!\left(k_{\max}\,K\left(p\bar{s} \;+\; v\,n\,\bar{s}^2\right)\right).
	\]
	In many high-dimensional sparse regimes, the $p\bar{s}$ term dominates the $v\,n\,\bar{s}^2$ term, but we retain both terms in the bound stated above.
	
	\textbf{3. Final model fitting:} After selection, FSCRE fits a robust MM-estimator to each of the $K$ sub-models. For a model of size $s_k$, the cost is $O(C_{\mathrm{iter}}\,n\,s_k^2)$, where $C_{\mathrm{iter}}$ denotes the number of reweighting iterations. Executed once per sub-model, the total cost of final fitting is $O(K\,n\,\bar{s}^2)$, which is independent of $p$ and typically negligible relative to the preprocessing and selection-loop costs.
	
	Combining the dominant terms yields the overall complexity bound reported in the main text.

	\section{Empirical Sensitivity Analysis for the Number of Models ($K$)}
	\label{sec:sensitivity}
	
	In Section~\ref{sec:method_ensemble}, we note that the number of models $K$ is a key tuning parameter controlling the complexity of the FSCRE ensemble. To assess the recommended range $K \in [5,10]$, we conduct a sensitivity analysis evaluating performance over a grid of $K$ values.
	
	We consider the \textit{Mixture Correlation} contamination scenario with $\text{SNR}=1.0$ and examine three sparsity levels ($50$, $100$, and $200$ active predictors out of $p=500$). For each configuration, we vary $K$ from $1$ to $20$ and compute the median MSPE, recall, and precision over 50 independent replications. For reference, we also report the single-model cellwise-robust baseline (CellRLARS), which corresponds to the special case $K=1$.
	
	\begin{figure}[h!]
		\centering
		\includegraphics[width=\textwidth]{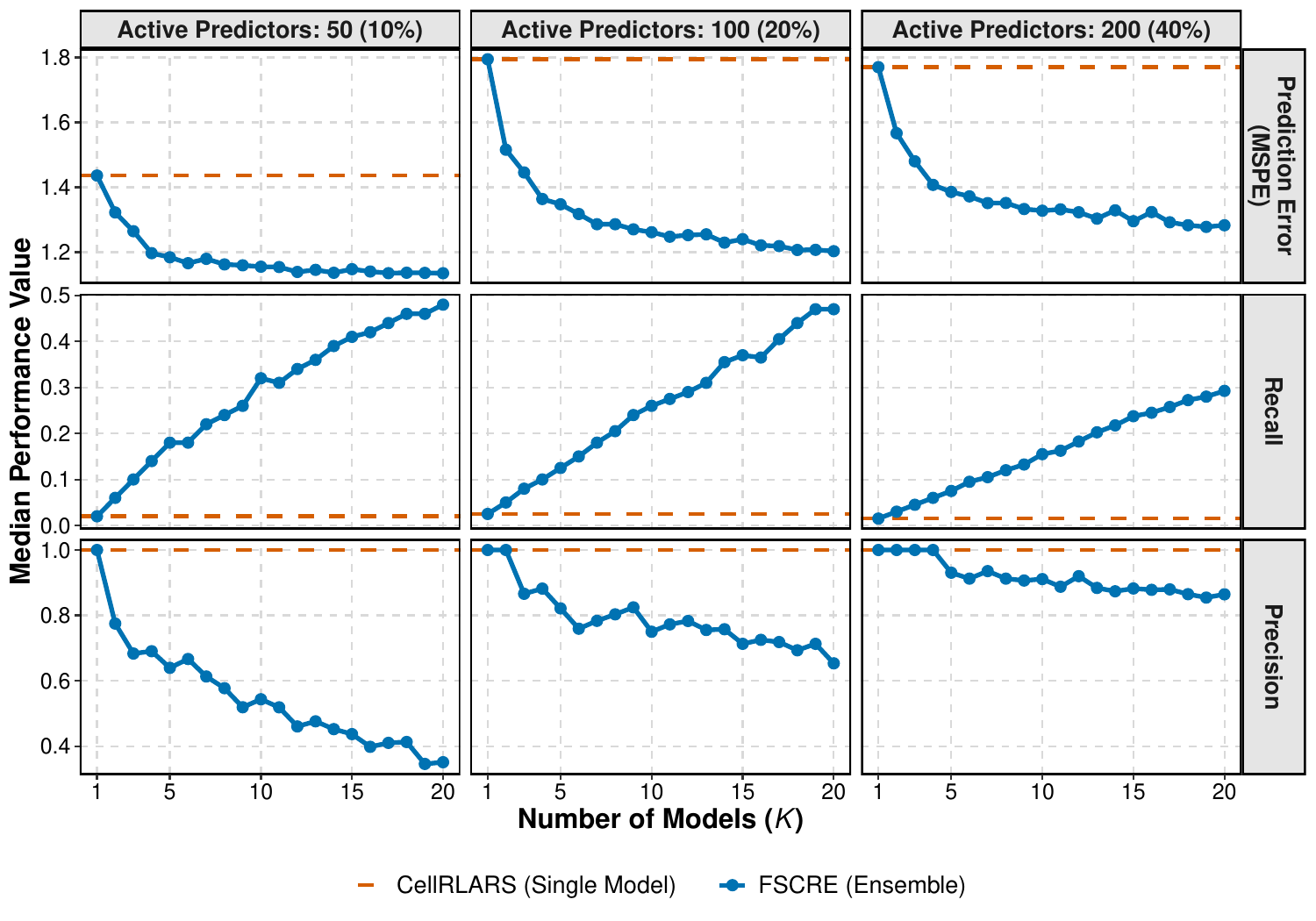}
		\caption{Median Mean Squared Prediction Error (MSPE), Recall, and Precision of the FSCRE algorithm as a function of the number of sub-models ($K$). Results are shown for the \textit{Mixture Correlation} scenario (SNR $= 1.0$) across three sparsity levels (50, 100, and 200 active predictors) over 50 replications.}
		\label{fig:K_sensitivity}
	\end{figure}
	
	The results, visualized in Figure~\ref{fig:K_sensitivity}, clearly illustrate the structural trade-offs governed by $K$:
	\begin{itemize}
		\item \textbf{Prediction Error (MSPE):} Across all sparsity levels, the prediction error exhibits a characteristic ``elbow.'' The MSPE drops sharply as $K$ increases beyond the single-model baseline (dashed line), reflecting the variance-reduction benefits of ensemble aggregation. Beyond $K=10$, the prediction error generally plateaus, indicating diminishing returns for out-of-sample accuracy.
		\item \textbf{Variable Selection (Recall and Precision):} Under severe, collinear contamination, a single cellwise-robust LARS path is excessively conservative and fragile, yielding near-zero recall. Increasing $K$ provides the ensemble with the necessary capacity to partition the predictor space and uncover masked signals, leading to a strict, nearly linear increase in Recall (middle row). However, this expanded capacity eventually leads to the inclusion of noise variables, reflected by the steady decline in Precision (bottom row).
	\end{itemize}
	
	The predictive arbitration stage of FSCRE is intended to balance this trade-off. In our experiments, the lowest MSPE is typically attained for $K \in [5,10]$, suggesting that this range is sufficient to capture the dominant predictive signals while limiting the accumulation of additional variables that may contribute little signal and increase variability. Accordingly, $K$ between 5 and 10 appears to provide a practical and computationally efficient default in the high-dimensional settings considered here.

	\section{Computational Environment and Software Details}\label{sec:supp_software}
	
	To ensure strict reproducibility of the simulation study and real-data application, all experiments are executed in a controlled computational environment. This section details the hardware specifications, system configurations, and exact software versions used to generate the results presented in the manuscript.
	
	\subsection{Computational Platform}
	All simulations and empirical timing evaluations are conducted on a Linux-based High-Performance Computing (HPC) cluster managed by the Slurm workload manager. The specific hardware and system configurations allocated for each array task are as follows:
	\begin{itemize}
		\item \textbf{Operating System / Environment:} Linux environment with GCC compiler version 14.2.0.
		\item \textbf{R Version:} R version 4.4.2.
		\item \textbf{Hardware Allocation:} 6 CPU cores and 12 GB of RAM per task. 
		\item \textbf{Threading Configuration:} To prevent resource contention between explicit parallelization (e.g., via the \texttt{parallel} and \texttt{foreach} packages) and implicit linear algebra multi-threading, all BLAS/LAPACK and OpenMP thread limits are strictly constrained to a single thread (\texttt{OMP\_NUM\_THREADS=1}, \texttt{MKL\_NUM\_THREADS=1}, \texttt{OPENBLAS\_NUM\_THREADS=1}).
	\end{itemize}
	
	\subsection{Software Dependencies and Package Versions}
	The proposed FSCRE methodology is implemented in the \texttt{srlars} R package (version 3.0.0, \citealp{srlars_package}). To guarantee consistency in the baseline comparisons and data preprocessing steps, the exact versions of all competitor algorithms and supporting utilities are explicitly enforced using the \texttt{remotes} package.
	
	\paragraph{Methodological and Competitor Packages:}
	\begin{itemize}
		\item \textbf{Data Cleaning \& Robust Correlations:} \texttt{cellWise} (v2.5.5; \citealp{cellWise_package}).
		\item \textbf{Competitor Ensembles:} \texttt{randomGLM} (v1.10-1; \citealp{song2013random}) for Random GLM; \texttt{glmnet} (v4.1-10; \citealp{glmnet_package}) for standard Elastic Net.
		\item \textbf{Casewise-Robust Baselines:} \texttt{pense} (v2.5.0; \citealp{pense_package}) for the PENSE algorithm; \texttt{robustHD} (v0.8.4; \citealp{robustHD_package}) for SparseLTS; \texttt{robustbase} (v0.99-6; \citealp{robustbase_package}) for standard robust MM-estimators.
		\item \textbf{Cellwise-Robust Baselines:} The CR-Lasso algorithm \citep{su2024cr} is run using the \texttt{regcell} package (accessed via GitHub). 
	\end{itemize}
	
	\paragraph{Bioinformatics Application (Bioconductor Packages):}
	The TCGA breast cancer (BRCA) proteogenomics dataset is retrieved and processed using the Bioconductor suite (Release 3.19 compatible). Key data-handling packages include:
	\begin{itemize}
		\item \texttt{curatedTCGAData} (v1.32.1; \citealp{curatedTCGAData_package})
		\item \texttt{TCGAutils} (v1.30.2, \citealp{TCGAutils_package})
		\item \texttt{MultiAssayExperiment} (v1.36.1, \citealp{MultiAssayExperiment_package})
		\item \texttt{SummarizedExperiment} (v1.40.0, \citealp{SummarizedExperiment_package})
	\end{itemize}
	
	\paragraph{Simulation Utilities:}
	Additional CRAN packages used for data simulation, parallel processing, and evaluation metrics include \texttt{MASS} (v7.3-65), \texttt{Matrix} (v1.7-4), \texttt{matrixStats} (v1.5.0), \texttt{doParallel} (v1.0.17), and \texttt{foreach} (v1.5.2). 
	
	A complete, automated environment setup script that installs these exact versions is provided in the public GitHub repository accompanying this manuscript: \url{https://github.com/AnthonyChristidis/FSCRE-Simulations}
	
	\bibliographystyle{apalike}
	\bibliography{FSCRE}
	
\end{document}